\documentclass[preprint2]{aastex}

\usepackage{graphicx}
\usepackage{graphics}
\usepackage{txfonts}
\usepackage{epsfig}
\usepackage{natbib}
\usepackage{times}
\usepackage{amssymb}
\usepackage{color}
\usepackage{longtable}

\newcommand\chan{{\it{Chandra}}}

\newcommand\numHMXB{79}
\newcommand\numOB{458}
\newcommand\numHMXBsun{54}
\newcommand\numOBsun{361}
\newcommand\numKM{133}

\shorttitle{Clustering between HMXBs and OB associations}
\shortauthors{Bodaghee et al.}

\begin{document}

\title{Clustering between high-mass X-ray binaries\\and OB associations in the Milky Way}

\author{A. Bodaghee and J. A. Tomsick}
\affil{Space Sciences Laboratory, 7 Gauss Way, University of California, Berkeley, CA 94720, USA}
\email{bodaghee@ssl.berkeley.edu}

\and

\author{J. Rodriguez}
\affil{Laboratoire AIM, CEA/IRFU - Universit\'e Paris Diderot - CNRS/INSU,\\ 
CEA DSM/IRFU/SAp, Centre de Saclay, F-91191 Gif-sur-Yvette, France}

\and

\author{J. B. James}
\affil{Dark Cosmology Centre, University of Copenhagen, Juliane Maries Vej 30, 2100 Copenhagen, Denmark\\ 
Astronomy Department, University of California, Berkeley, CA 94720, USA}

\begin{abstract}
We present the first direct measurement of the spatial cross-correlation function of high-mass X-ray binaries (HMXBs) and active OB star-forming complexes in the Milky Way. This result relied on a sample containing \numHMXB\ hard X-ray selected HMXBs and \numOB\ OB associations. Clustering between the two populations is detected with a significance above 7$\sigma$ for distances $<$1\,kpc. Thus, HMXBs closely trace the underlying distribution of the massive star-forming regions that are expected to produce the progenitor stars of HMXBs. The average offset of 0.4$\pm$0.2\,kpc between HMXBs and OB associations is consistent with being due to natal kicks at velocities of the order of 100$\pm$50\,km\,s$^{-1}$. The characteristic scale of the correlation function suggests an average kinematical age (since the supernova phase) of $\sim$4\,Myr for the HMXB population. Despite being derived from a global view of our Galaxy, these signatures of HMXB evolution are consistent with theoretical expectations as well as observations of individual objects.

\end{abstract}


\section{Introduction}

High-mass X-ray binaries (HMXBs) are systems in which a compact object (usually a neutron star, but sometimes a black hole candidate) accretes from a massive stellar companion ($M\gtrsim10$\,$M_{\odot}$). The mass-age relation in stellar evolution predicts that $\sim10^{7}$\,yr elapses between starbirth and supernova in high-mass stars \citep{Sch92}. Thus, HMXBs are relatively young systems which are not expected to migrate far from their birthplaces: sites with a recent history of massive star formation, i.e., the OB associations that trace the Galactic spiral arms \citep[e.g.][]{Bro99}.

\begin{figure}[!t] \centering
\includegraphics[width=8cm,angle=0]{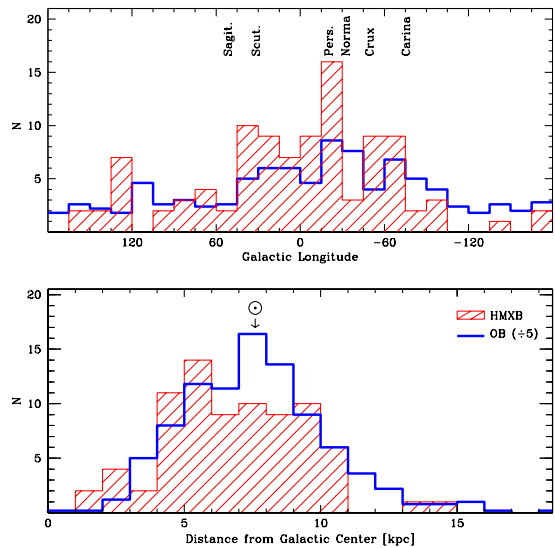}
\caption{Distribution of Galactic longitudes and galactocentric distances (when known) of HMXBs (shaded histogram) and OB associations from \citet{Rus03} (thick curve: divided by 5). The locations of the spiral arm tangents are shown. The distributions of HMXBs in these projected spatial directions are compatible with those of OB associations.}
\label{long_dist}
\end{figure}

Observational evidence linking Galactic HMXBs and OB associations has been demonstrated in individual cases in which the connection is attested by other factors: i.e., consistent proper motions and distances, the position of the HMXB donor star with respect to the main sequence (MS) of the OB association on color-magnitude diagrams, etc. On a Galactic scale, however, the evidence is limited to comparing distributions of Galactic longitudes ($l$) or galactocentric distances ($R$) \citep[e.g.,][]{2002AaA...391..923G,Dea05,Lut05,Bod07,Lut08}. Both HMXBs and OB associations show maxima around Galactic longitude $l\sim30^{\circ}$ which corresponds roughly to the direction of the Norma and Inner Perseus Arms (upper panel of Fig.\,\ref{long_dist}). Moderate peaks are found towards the Carina, Scutum, and Sagittarius Arms. This suggests that HMXBs are spatially correlated with active sites of massive-star formation. This correlation is further exemplified by the similarity in the distributions of galactocentric distances of HMXBs and OB associations: both distributions can be approximated by Gaussians centered on the galactocentric distance of the Sun (lower panel of Fig.\,\ref{long_dist}). 

\begin{figure*}[!t] \centering
\includegraphics[width=16cm,angle=0]{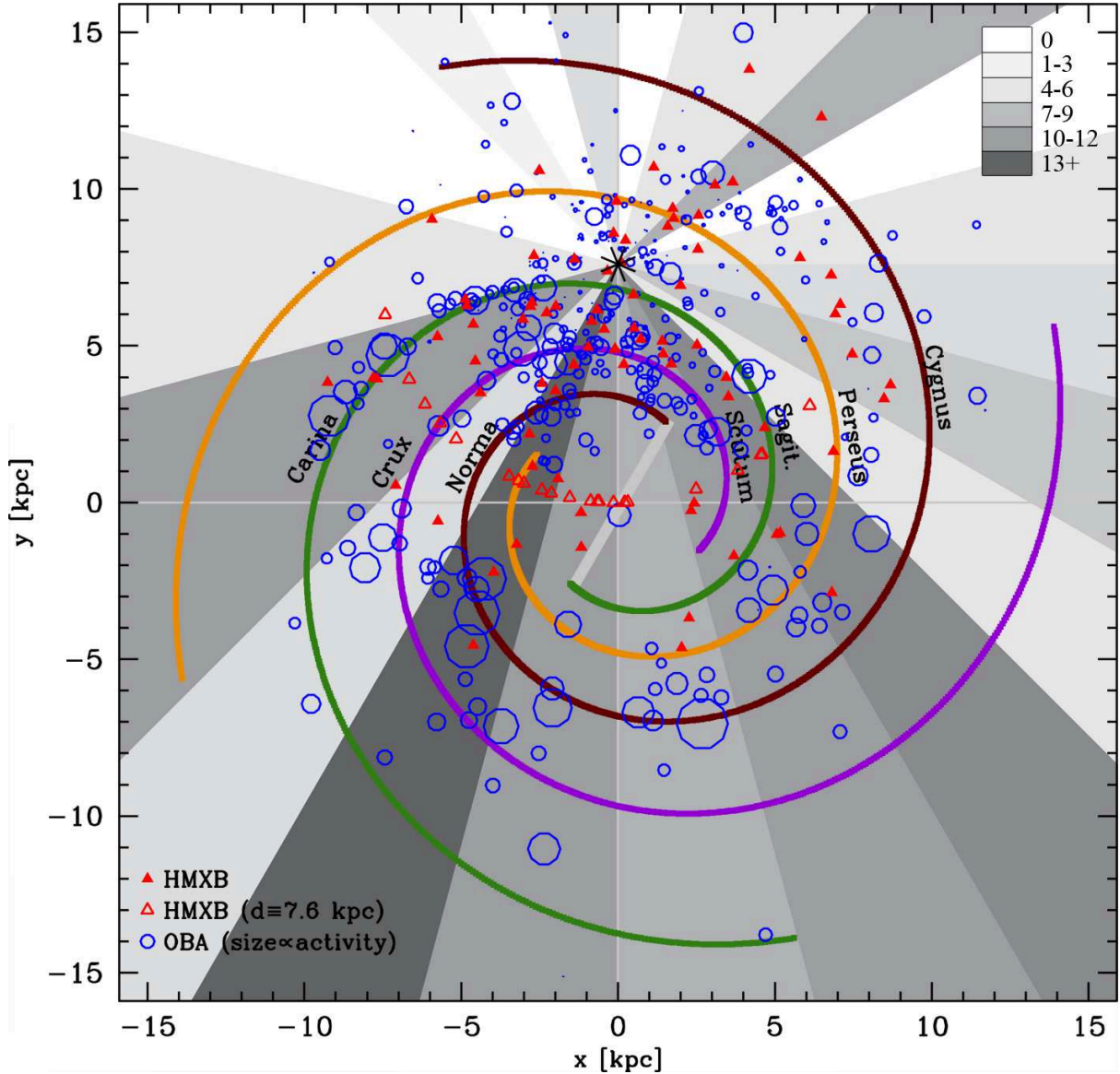}
\caption{Galactic distribution of HMXBs whose distances are known (\numHMXB, filled triangles)
and the locations of OB associations from \citet{Rus03} (\numOB, circles). The symbol size of the latter is proportional to the amount of activity in the association ($\equiv$ amount of ionizing photons as determined from the radio continuum flux). The spiral arm model of \citet{Val08} is overlaid with the Sun situated at 7.6 kpc from the Galactic Center (GC). HMXBs whose distances are not known have been placed at 7.6 kpc (23, empty triangles), i.e., the Sun-GC distance assumed in \citet{Val08}; note that these sources are not included in the cross-correlation analysis. The shaded histogram represents the number of HMXBs in each 15$^{\circ}$ bin of galactic longitude as viewed from the Sun. }
\label{gal_disp_hmxb}
\end{figure*}

Recent efforts to describe the Galactic distribution of HMXBs raised the intriguing possibility of an offset between the peak in the HMXB longitudinal distribution with respect to the tangent to the Norma Arm \citep{Lut05,Dea05,Bod07,Lut08}. This direction features the highest formation rate of massive stars \citep{Bro00}. An offset would imply that we are witnessing the systematic delay ($\equiv$ kinematical age) between the epoch of star formation and the moment when these massive binaries became X-ray emitters. Figure\,\ref{gal_disp_hmxb} presents the population of HMXBs and OB associations in the Milky Way as it would appear to an observer situated above the Galactic Plane. Thanks to an increase in HMXB statistics, and by using a recent Galactic spiral arm model \citep{Val08}, we can now attribute the apparent offset, at least partly, to foreground sources that are unrelated to the Norma Arm, and to sources located towards the tangent to the Inner Perseus Arm. The extension of the latter arm into the Inner Galaxy was not considered in previous spiral arm models \citep[see also][]{Hou09}. 

Clearly, a precise study of the spatial relation of massive star-forming regions and the HMXBs they spawn can offer valuable insight into stellar and Galactic evolution. The drawback with longitudinal or galactocentric distance distributions is that these histograms consider only a single dimension at a time. In effect, they project an entire spatial component ($R$ in the case of a longitudinal distribution, and $l$ in the case of a galactocentric distance distribution) onto the other component. Precious information about the true proximity of objects in the two populations is lost. Hence, these methods can not accurately assess the spatial relation, restricting their ability to provide the characteristic scales that would help us understand how HMXBs are distributed in the Milky Way. 

A better approach is to consider the spatial (or cross-) correlation function $\xi$ (or SCF). This function provides a statistical measure of the clustering of point sources in multi-dimensional maps of the sky. It is frequently used to relate active galactic nuclei with clusters of galaxies \citep[e.g.][]{Pee80,Lan93}. In Section\,\ref{xi}, we adapt the methods described in these papers to suit the study of HMXBs on a Galactic scale, and we assess biases in Section\,\ref{bias}. Our findings are discussed in the context of stellar and Galactic evolution in Section\,\ref{disc}.

%

\section{Data}
\label{data}

We selected all sources from \citet{Bod07} which are confirmed or strongly suspected of being HMXBs. This catalog contains hard X-ray detected sources ($\gtrsim$20\,keV), meaning that their detection does not depend on the photoelectric absorption that blocks soft X-rays ($\lesssim$5\,keV). A number of discoveries have been made since the original catalog was published, and these new sources were included as well. Many HMXBs are listed on the IGR Sources Page which we keep current\footnote{\texttt{http://irfu.cea.fr/Sap/IGR-Sources}}. The date of April 1, 2011, represents the cut-off after which we no longer considered new information.

For each HMXB, we searched through the literature and extracted the most recent distance measurement and related uncertainty when available. We strove to consider only distances that were derived from observations of the optical/IR counterpart, and not distances assumed from, e.g., being in the line of sight of a particular spiral arm or the Galactic Center (GC). Any HMXB suspected of or confirmed to belong to the Magellanic Clouds was excluded. The final sample consists of the \numHMXB\ X-ray selected HMXBs listed in Table\,\ref{tab_hmxb}. The positions and distances of OB-forming complexes are drawn from \citet{Rus03}. These complexes are large and excited H\,II regions which include OB associations, but we will use the term ``complex'' interchangeably with ``association.'' We rejected OB associations that are located more than 20\,kpc from the GC resulting in \numOB\ associations.

We emphasize that the HMXB and OB samples are drawn exclusively from the Milky Way. \citet{Lin10} argue that galaxies with a metallicity comparable to that of the Milky Way produce few luminous black hole HMXBs such as \object{Cyg\,X-1} or \object{SS\,433} \citep[see also][]{Mir11}. Hence, the HMXB sample contains mostly neutron star HMXBs classified into one of three groups: the transient emitters with Be companions (BEXBs), the mostly persistent emitters with supergiant companions (SGXBs), and the fast X-ray transients with supergiant companions (SFXTs).

\begin{figure}[!ht] \centering
\includegraphics[width=8cm,angle=0]{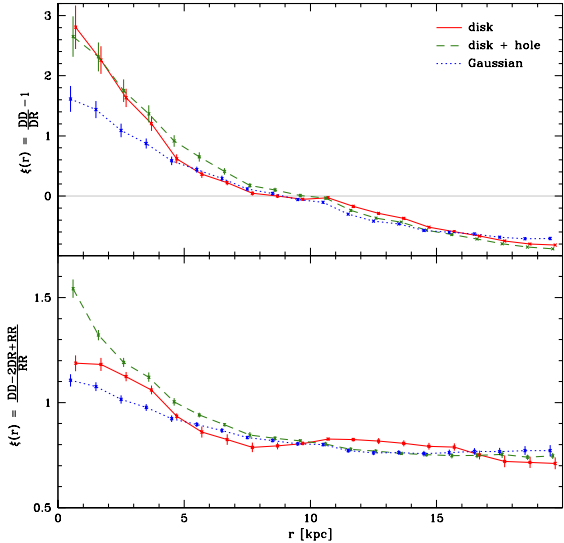}
\caption{Spatial correlation functions $\xi (r)$, where $r$ represents the distance from a given HMXB (in kpc), are presented for both estimators. Three random distributions were considered: a Gaussian ring centered at 7.6\,kpc with $\sigma_{z} = 2.5$\,kpc; a uniform disk for $0 \leq r \leq 16$\,kpc; and a uniform disk with a hole at 2\,kpc from the GC. The curves are slightly shifted horizontally in order to allow the 1-$\sigma$ error bars to be clearly distinguished. At short distances from a given HMXB, $\xi > 1$. This suggests that the closest neighbors of an HMXB tend to be observed OB associations rather than OB associations drawn from a random distribution.}
\label{xi_all}
\end{figure}

\section{The Spatial Correlation Function}
\label{xi}

For a given HMXB in a volume element $\delta V_{1}$, the probability $\delta P$ (in excess of Poisson) of finding an OB association in a volume element $\delta V_{2}$ separated by a distance $r$ (in kpc) is:

\begin{equation}
\delta P = n_{1}n_{2}  \left [ 1 + \xi (r)  \right ] \delta V_{1}\delta V_{2}
\label{eq_prob}
\end{equation}

The mean number densities are given by $n_{1}$ and $n_{2}$ where the subscripts refer to the two populations: HMXBs (1) and OB associations (2). These are young objects that remain within $45\pm20$ pc of the Galactic Plane \citep{Ree00}, so we assume that the displacement along the $z$-axis is negligible, and surface (rather than volume) elements are used.

\begin{figure}[!t] \centering
\includegraphics[width=8cm,angle=0]{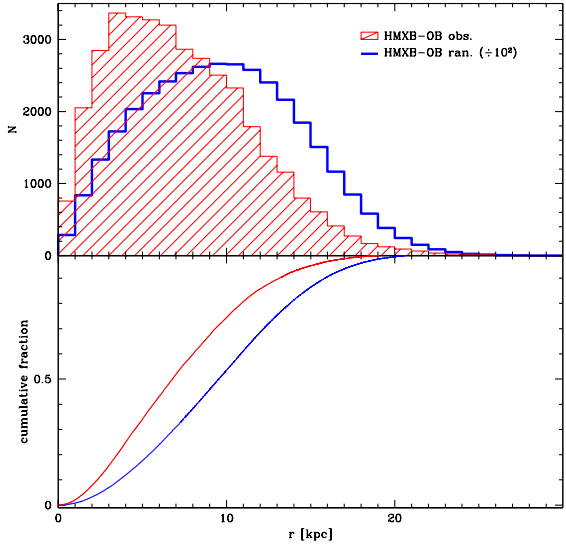}
\caption{Distributions of distances between pairs of HMXBs and observed OB associations (red shaded histogram), and between pairs of HMXBs and Gaussian-randomized OB associations (divided by $10^{2}$, blue histogram). A KS-test (lower panel) yields a probability $<10^{-7}$ of statistical compatibility. }
\label{dist_hmxb_OB}
\end{figure}

Rings of 1-kpc thickness (from 0 to 20\,kpc) are constructed around each HMXB. In each ring, we count the number of observed OB associations, and we keep a separate tally of the number of OB associations drawn from a random distribution. An HMXB-OB pair is referred to as a $DD_{12}$-pair (for data-data pair), while an HMXB-random OB pair is labelled $DR_{12}$ (for data-random). In this way, we construct $\xi$ for each radius according to the definition of \citet{Pee80}:

\begin{equation}
\xi (r) = \frac{ n_{R} DD_{12} }{ n_{D} DR_{12} } -1
\label{xi_pee80}
\end{equation}

If $\xi = 0$, which implies that each ring contains as many $DD$-pairs as $DR$-pairs, then Eq.\,\ref{eq_prob} is simply a uniform Poissonian probability. However, if $\xi > 0$, then there is a higher chance of an HMXB having a neighbor that is an observed (rather than a randomized) OB association. \citet{Lan93} propose a more robust estimator for $\xi$ in which the variance is nearly Poissonian:

\begin{equation}
\xi (r) = \frac{DD_{12} - DR_{12} - DR_{21} + RR_{12}}{RR_{12}}
\label{eq_xi_pee80}
\end{equation}

Three random distributions were considered: a disk model in which the OB associations were uniformly distributed between 0 an 16\,kpc from the GC; a disk model with a hole, i.e., a uniform distribution from 2\,kpc to 16\,kpc from the GC; and a Gaussian distribution centered at 7.6\,kpc from the GC with $\sigma_{z} = 2.5$\,kpc. The boundaries, centroid, and width of the Gaussian were chosen to be consistent with the observed distribution of HMXBs and OB associations (see Fig.\,\ref{long_dist}). Each random distribution contained the same number of OB associations as the observed OB sample (\numOB) occupying similar surface areas so $n_{D}$ and $n_{R}$ are equal. One thousand trials were performed with each randomization model. The average $\xi$, along with its 1-$\sigma$ uncertainties, are shown in Fig.\,\ref{xi_all}.

All three randomization models and both estimators of $\xi$ lead to the same result: near an HMXB, the probability of finding a known OB association is higher than expected from Poisson statistics. In other words, HMXBs and OB associations are clustered together. Our measured value of $\xi(r<1$\,kpc$)=1.61\pm0.22$ (Gaussian case) implies that $xDD \sim DR$ (where $x\sim2.6$), i.e.,  around a given HMXB, one counts between 2 and 3 times as many observed OB associations as randomized ones. We are well below the value of $\sim$\numOB\ anticipated for perfectly-correlated samples, but statistically well above the value expected for no correlation whatsoever ($\xi\sim0$).

For $r < 3$\,kpc, the clustering signal from the \citet{Pee80} estimator is 7--11$\sigma$ in excess of Poisson (up to 17$\sigma$ for the \citet{Lan93} estimator). The observed and random surface-density distributions are statistically compatible only at large radii (i.e. $r \gtrsim 5$\,kpc). Notice that while $\xi (r<1\mathrm{\,kpc}) > \xi (2\leq r<3\mathrm{\,kpc})$, the standard deviation of $\xi$ decreases as we move away from an HMXB (up to a certain limit). This is simply a consequence of the increase in the surface areas, and hence the number of possible pairs, with increasing radius. 

Of the three models, the Gaussian distribution provides the best approximation since its $\xi$ deviates the least from Poisson statistics at short distances. The disk distributions (with and without a central hole) are no longer considered. 

\begin{figure}[!t] \centering
\includegraphics[width=8cm,angle=0]{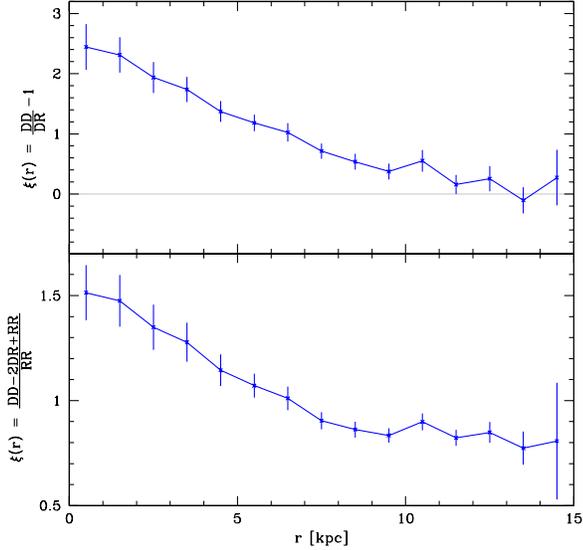}
\caption{Same as Fig.\,\ref{xi_all} but restricted to objects within 8\,kpc from the Sun. The random catalog assumes a Gaussian-ring distribution. Clustering between \numHMXBsun\ HMXBs and \numOBsun\ OB associations persists with a statistical significance of 7$\sigma$. }
\label{xi_hmxb_8kpc}
\end{figure}

\section{Statistical and Systematic Biases}
\label{bias}

Observational biases affect the correlation function of the objects in the survey. This is not trivial as the distribution of HMXBs and OB associations are, \emph{a priori}, not well understood. The type of random distribution that one chooses leads to a systematic bias. This is readily apparent from Fig.\,\ref{dist_hmxb_OB} where the distances between pairs of HMXBs and observed OB associations are compared with the distances between pairs of HMXBs and Gaussian-randomized OB associations. Despite the fact that the distributions have similar mean and variance values, the higher moments of the distributions (skewness and kurtosis) disagree. A Kolmogorov-Smirnov test yields a probability of less than $10^{-7}$ of statistical compatibility between the two distributions. 

Selection bias is also important. An HMXB situated far from the Sun needs to be correspondingly more luminous in order to be detected in the X-rays. Large distances also make the optical classification as an HMXB more difficult: line spectroscopy of the donor star is hindered by reddening and absorption from interstellar material in the Galactic Disk. This leads to a preponderance of HMXBs (as well as observed OB associations) close to the Sun, and a paucity of such objects situated behind the GC and its bar: a region referred to as the ``zona Galactica incognita'' \citep[e.g.,][]{Val02}. The random distributions of HMXBs and OB associations that we generated did not consider this observational sampling bias. 

\begin{figure}[!t] \centering
\includegraphics[width=8cm,angle=0]{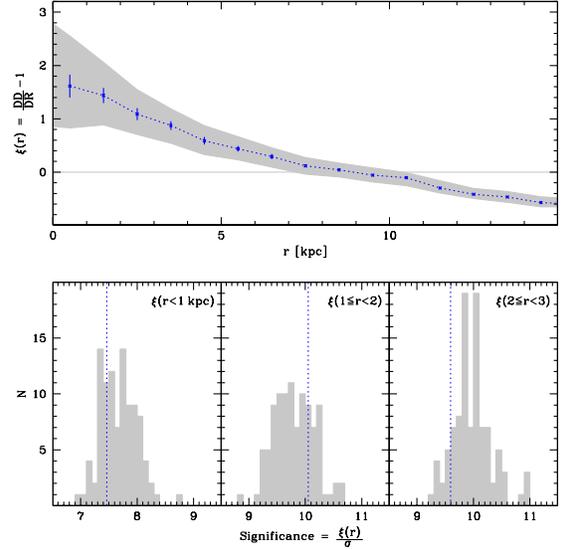}
\caption{The clustering remains significant even when the HMXBs are perturbed within a range defined by the uncertainties on their distances. \emph{Top}: Three-sigma boundaries (gray shaded region) from multiple SCFs generated by randomly shuffling (according to a Gaussian distribution) the line-of-sight distance to each HMXB within its distance uncertainty. The blue dotted line and data points represent the non-perturbed (i.e., observed) SCF as presented in Fig.\,\ref{xi_all}. \emph{Bottom}: The distribution of the clustering significance from all reshuffling trials where significance is defined as $\xi (r) / \sigma$. From left to right, the panels present the distributions for $\xi (r<1 \mathrm{\,kpc})$, $\xi (1\leq r<2)$, and $\xi (2\leq r<3)$.  }
\label{xi_hmxb_rand_dist}
\end{figure}

It is reasonable to expect symmetry in the distribution of HMXBs and OB associations around the Galactic Disk. However, the fact that there are less HMXBs detected (with distances measured) in the zona incognita will surely affect the SCF. Therefore, we generated the SCF for a restricted boundary corresponding to a circle of 8-kpc radius around the Sun---a perimeter inside of which most HMXBs and OB associations should be detectable and their distances known with reasonable accuracy. When only pair counts of objects (\numHMXBsun\ HMXBs and \numOBsun\ OB associations) in the Solar neighborhood are considered, the clustering signal at small radii persists with a statistical significance of 7$\sigma$ for $r < 1$\,kpc (see Fig.\,\ref{xi_hmxb_8kpc}). 

Some HMXBs have distances with uncertainties that are on the order of, or larger than, the 1-kpc spatial binning that we used. The counts per ring will fluctuate depending on the distance precision quoted in the literature. To account for this bias, we shuffled each HMXB according to a Gaussian profile within its line-of-sight distance range as defined by its error bars. When an error bar was not provided in the literature, we set the uncertainty range to be equal to the average of the known error bars of the sample (1.5\,kpc). Then, we regenerated $\xi$ for each of the 100 randomly-shuffled HMXB distributions. Keep in mind that the OB distribution is itself randomized $10^{3}$ times per trial, which means that the total number of different HMXB-OB configurations that we tested is actually $10^{5}$. 

Even when the HMXBs are perturbed within their line-of-sight uncertainties, the clustering signal remains above 7$\sigma$ in all trials for $r<3$\,kpc from a given HMXB (Fig.\,\ref{xi_hmxb_rand_dist}). Consider a conservative assumption that says the clustering is significant only when $\xi \ge1$ (i.e., when each ring around an HMXB contains at least twice as many observed OB associations as random OB associations). We find that 97\% of the correlation functions describing the shuffled distributions satisfy this stringent criterion. A little over half of the trials (53\%) yield a $\xi$ value that is above this cutoff at the 3-$\sigma$ level. Based on the $>$99\% confidence interval displayed in Fig.\,\ref{xi_hmxb_rand_dist}, we estimate a systematic uncertainty of 5\% on $\xi$.

\begin{figure}[!t] \centering
\includegraphics[width=8cm,angle=0]{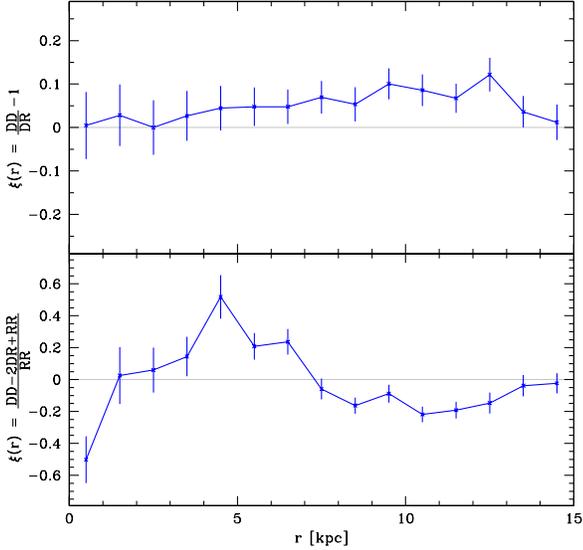}
\caption{The SCF is presented comparing the locations of OB associations, and an equivalent number of points representing the four-arm spiral model of \citet{Rus03}. The randomized catalog assumes a Gaussian ring distribution and the Galactocentric distance of the Sun is set to 8.5\,kpc, i.e., the value used in \citet{Rus03}.}
\label{xi_OB_arms}
\end{figure}

Potential differences in the scale heights of the two populations has thus far been ignored. A few HMXBs present high velocities tangential to the Galactic Plane \citep[e.g., \object{4U~1907$+$09},][]{Gva11}. By projecting this velocity component onto the Galactic Plane, we can expect to underestimate the true offset between an HMXB and a nearby OB association. This will lead to samples that are more correlated than they should be. Therefore, we calculated the Galactic scale height for members of our HMXB and OB samples. The average scale heights are in good agreement at 0.1\,kpc for both samples with standard deviations of 0.1 and 0.4 for HMXBs and OBs, respectively. This suggests that at the 3-$\sigma$ level, HMXBs are located at most 0.4\,kpc above or below the Plane. This value is consistent with the characteristic scale and average minimum offset that are discussed in Section\,\ref{disc}. Ninety percent of the HMXBs (70/79) are within 0.25\,kpc of the Plane with no HMXBs located further than 0.55\,kpc from the Plane. Therefore, we are justified in using the 2-D distribution since any increase in the correlation significance caused by this projection effect will be negligible given that these offsets are less than the 1-kpc width of each ring in our correlation analysis.

\begin{figure}[!t] \centering
\includegraphics[width=8cm,angle=0]{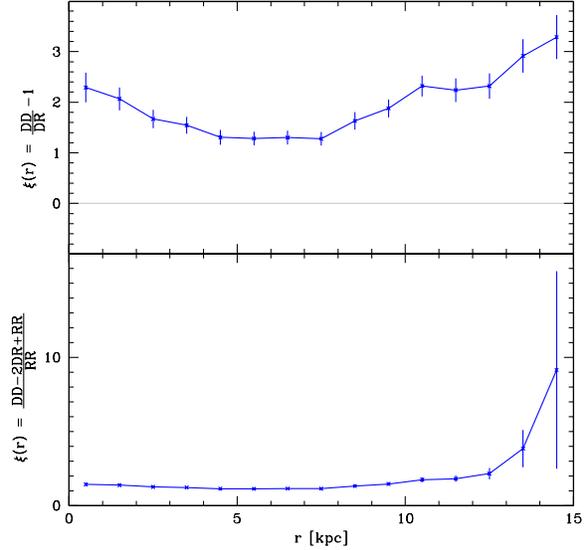}
\caption{The SCF is presented for the case in which a randomized (Gaussian ring) HMXB catalog is taken as the observed set and is then compared with the OB associations. }
\label{xi_hmxb_rand}
\end{figure}

We tested the correlation between HMXBs and OB associations against the spiral arm model of \citet{Rus03} whose equations represent the best-fitting four-arm logarithmic spiral to the locations of the star-forming complexes shown as blue circles in Fig.\,\ref{gal_disp_hmxb}. Here, we adopted the Sun-GC distance of 8.5\,kpc assumed in the model. This model yields an overdensity of points describing an arm in the inner Galaxy compared with the same arm in the outer Galaxy (i.e., as one loops through the range of angles that trace an arm from the inner to the outer Galaxy, the actual distance separating successive points becomes larger). Thus, in order to reduce this bias, we extracted 115 points chosen at random along each of the four arms (460 total: i.e. equivalent in number to the \numOB\ observed OB associations). We performed a visual check and verified that the spiral pattern was easily discernible with no major gaps ($>$ 2\,kpc) along an arm.

\begin{figure}[!t] \centering
\includegraphics[width=8cm,angle=0]{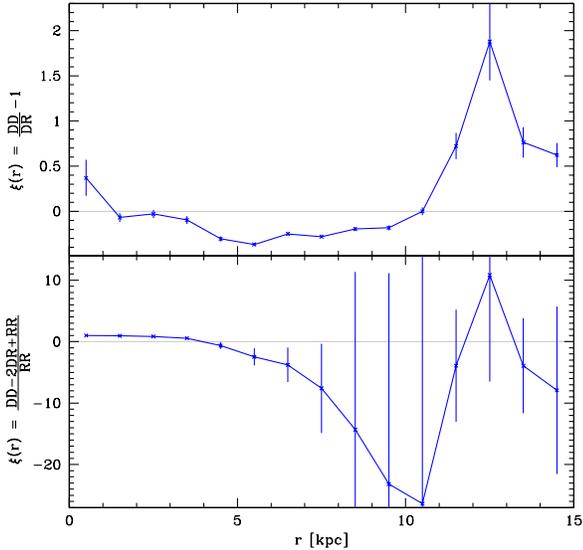}
\caption{The SCF compares HMXBs with \numKM\ globular clusters from \citet{Bic06} that are located within 20\,kpc of the GC. The random catalog of globular clusters assumes an exponential decay law. The SCF is consistent with Poisson statistics at close radii, which means that the neighbors of HMXBs are just as likely to be drawn from the observed distribution as they are from a random one: i.e., there is no clustering between HMXBs and globular clusters. }
\label{xi_lmxb}
\end{figure}

Figure\,\ref{xi_OB_arms} shows that there is no significant correlation between OB associations and the spiral arms with $\xi(r) \sim 0$ for all $r$ (i.e., $DD \sim DR$ for all $r$). Similarly, we found no significant clustering between HMXBs and the points representing the spiral arms. Restricting the analysis to objects in the Solar vicinity (i.e., within 8\,kpc of the Sun) leads to similar conclusions, so we can rule out an apparent lack of distant OB associations (or HMXBs) skewing the results. In other words, around any given OB association (or HMXB), one tends to find as many neighbors drawn from a randomized Gaussian ring distribution, as neighbors representing the spiral arms. These comparisons of our observed stellar samples with the Galactic models allow us to conclude, as acknowledged by \citet{Rus03}, that such spiral arm models are an overly simplistic representation of the real picture.

In another test, we randomized the HMXB distribution (using the Gaussian ring approximation) and we assumed that this sample was the observed HMXB set. The $\xi$ relating randomized HMXBs and OB associations is presented in Fig.\,\ref{xi_hmxb_rand}. There is a moderate deviation from Poisson statistics at short radii for the \citet{Pee80} estimator (whose values are $>$0 at all radii, as expected since the distributions are similar), while there is no deviation for the \citet{Lan93} estimator. This suggests that even though the randomization model mimics the global distribution of the observed HMXB sample, it is the \emph{observed} HMXB distribution that is clustered with the OB associations. 

Clustering between HMXBs and OB associations is expected. How would $\xi$ react if HMXBs were compared to a population of sources for which no spatial correlation is expected? We tested the correlation functions of HMXBs against a set of \numKM\ globular clusters from \citet{Bic06} that are located less than 20\,kpc from the GC. Globular clusters contain older populations such as cool KM dwarf stars and low-mass X-ray binaries (LMXBs). Unlike HMXBs, globular clusters (and LMXBs) are densely packed in the Galactic Bulge and their numbers drop exponentially with increasing radius from the GC \citep{Bod07}. The random distribution of globular clusters was thus modeled as an exponential decay law adjusted to be consistent with the observed distribution. Figure\,\ref{xi_lmxb} shows that $\xi$ between HMXBs and globular clusters is consistent with 0 for $r < 1$\,kpc from an HMXB. For any given HMXB, its immediate neighbors were just as likely to be drawn from the observed set as from the random set. Shot noise dominates at large distances. 

The results of these tests suggest that the cross-correlation function is assessing the real spatial clustering of populations (within the uncertainties) rather than some global property that they share. In other words, the significant clustering that we measured between HMXBs and OB associations (Fig.\,\ref{xi_all}) is due essentially to the physical locations of the individual objects in space, as opposed to being an attribute of the overall populations, e.g., that they appear to trace the spiral arms, or that they seem to follow a Gaussian distribution, etc.

\begin{figure}[!t] \centering
\includegraphics[width=8cm,angle=0]{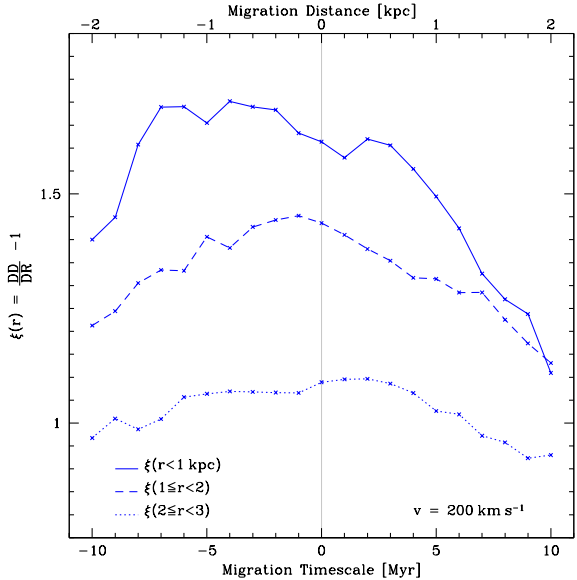}
\caption{The values of the SCF that result from shifting the positions of the HMXBs along their Galactic orbits. The shifts correspond to 1--10\,Myr (in steps of 1\,Myr) of circular motion at a velocity of 200\,km\,s$^{-1}$ in the forward ($+$) and reverse ($-$) directions of time. The continuous curve represents the values for $\xi (r<1 \mathrm{\,kpc})$, the dashed curve is $\xi (1\leq r<2)$, and the dotted curve is $\xi (2\leq r<3)$.}
\label{xi_hmxb_rot}
\end{figure}

\section{Discussion}
\label{disc}
There are several questions that we seek to address using the SCF. What is the characteristic scale of the clustering between HMXBs and OB associations? Can this scale be used to constrain the amount of migration due to the effects of Galactic rotation \citep[see, e.g.,][]{Ram94}, and perhaps due to the primordial ``kick'' velocity imparted to the system during the formation of the compact object? Overall, what can this scale tell us about the evolutionary history of HMXBs? 

\subsection{Galactic Rotation}
\label{rot}

According to the density-wave theory of spiral galaxies \citep{Lin64,Lin69}, the velocity of the spiral pattern is only half that of the material moving in and out of the arms. These arms appear stationary to an observer in a non-inertial reference frame that is rotating at the pattern speed, while the motions of individual objects within the Galaxy are not. As material encounters the quasistatic density waves, it is shocked and compressed which allows the Jeans criterion to be satisfied locally, triggering fragmentation of the cloud, and ultimately star formation. 

Massive stars in binary systems have a main sequence (MS) lifetime of around $5\times10^{6}$\,yr \citep{Sch92}. The exact age depends strongly on the stellar mass and spectral type. In addition, there is a systematic delay of $\sim4\times10^{6}$\,yr between the supernova phase and the X-ray emission phase \citep{Van83}. The X-ray phase lasts only $10^{4}$\,yr \citep{Ibe95} which is negligible on these timescales. In 10\,Myr, a combination of global (Galactic rotation) and local effects (e.g., supernova kicks, cluster outflows, and dynamical ejections) will cause newly-formed O and B stars to eventually migrate away from the dense gaseous regions in which they were born. In principle, each spiral arm includes three components \citep{Lin69}: a dusty strip in which the gas concentration and OB star formation rate are highest; a lane of luminous and recently-formed OB stars that have ionized large regions of gas (H\,II); and a wider lane of post-MS, red, or dying supergiant stars with smaller H\,II regions. The HMXBs should be found among the recently-formed OB stars and the post-MS stars.

Outside of the inner Galaxy (i.e., galactocentric radius $R\gtrsim2$\,kpc), objects orbit around the GC at a velocity of $\sim$200\,km\,s$^{-1}$ \citep[e.g.,][]{Mer92,Bra93,Glu99,Sof09}. Thus, a star at the Solar galactocentric radius (7.6\,kpc) will have moved $6.3\times10^{15}$\,km (200 pc) in 1\,Myr (defined as the migration timescale or $\tau$), assuming a circular orbit. In this manner, we rotated the HMXBs in space along their galactocentric orbits corresponding to $\tau=$ 1--10\,Myr (in steps of 1\,Myr) of circular motion in both the forward ($+$) and reverse ($-$) directions of time (clockwise and counter-clockwise, respectively, in Fig.\,\ref{gal_disp_hmxb}). Figure\,\ref{xi_hmxb_rot} presents the values of the SCF resulting from these shifts for small radii ($r < 3$\,kpc from an HMXB). When the radius is larger than $\sim$5\,kpc, $\xi$ is compatible with Poissonian statistics, and so these curves are omitted for clarity. 

Certain shifts increase the amplitude of the clustering signal. Increases in the amplitude of $\xi$ are directly related to an increase in the number of observed OB associations contained within the specified radius around the shifted HMXBs. This is because the randomized OB distribution is axisymmetric with respect to the GC, and so the number of random OB associations inside the radius remains relatively constant under shifts in either direction. Given that these shifts are small ($\lesssim 2$\,kpc), their effect on $\xi$ diminishes with increasing distance from an HMXB, as illustrated by the trend towards flatter curves for $\xi (1 \leq r < 2 \mathrm{\,kpc})$ and $\xi (2\leq r<3 \mathrm{\,kpc})$. We are mainly interested in clustering at short scales so, henceforth, our discussion will focus on $\xi (r<1 \mathrm{\,kpc})$ (the solid curve in Fig.\,\ref{xi_hmxb_rot}). At small radii, the \citet{Pee80} estimator provides sufficiently reliable results while also being more efficient computationally compared with the \citet{Lan93} estimator. 

If HMXB migration were related solely to Galactic rotation, then we would expect to see the amplitude of $\xi$ maximized for migration timescales around $-$10\,Myr. Instead, we find that the amplitude of $\xi$ is maximized for migration timescales between $-2$ and $-7$\,Myr. This suggests that young binary systems (prior to the formation of the compact object) are gravitationally bound to their birthplaces, in most cases. 

Another consequence of Galactic rotation is that more HMXBs should appear on the leading edge of a spiral arm than on its trailing edge (because of the factor 2 difference between the bulk and pattern speeds). Visual inspection of the Carina Arm (Fig.\,\ref{gal_disp_hmxb}) reveals no clear asymmetry in the distribution of HMXBs with equivalent numbers on either side of the arm. In the inner Galaxy, the tight spacing between arms combined with the uncertainty in the radial direction prevents us from visually assigning an HMXB to a specific arm. Nevertheless, since $\xi$ is a good indicator of the average proximity of HMXBs to the OB associations that trace these arms, we should see an asymmetry in its distribution with higher values of $\xi$ at negative timescales than at positive timescales. 

Indeed, there is a slight preference for migration towards the leading edge of spiral arms as hinted at by the asymmetric distribution and the downward trend in $\xi$ for $-5$\,Myr $\lesssim \tau \lesssim +5$\,Myr. The amplitude of $\xi$ is greater for negative migration timescales ($\tau \sim -4$\,Myr) than it is at rest ($\tau=0$\,Myr). However, the amplitude of $\xi$ for small positive migration timescales (i.e., $+2$\,Myr $\lesssim \tau \lesssim +3$\,Myr) is equivalent to that at rest which suggests that some HMXBs migrate in directions different from that of the bulk motion. This is another indication that we are not witnessing the effects of Galactic rotation alone.

\subsection{Kick Velocities}
\label{kick}

Galactic rotation can not fully explain the behavior of $\xi$ under different migration timescales. One mechanism suspected of giving an HMXB a significant velocity is the primordial kick inherited by the binary during the supernova event that created the compact object. This runaway velocity (with respect to its OB association) can be produced by asymmetric supernovae \citep{Shk70}, or from recoil due to anisotropic mass loss from the primary to the secondary \citep{Bla61}. Factors external to the binary can also allow it to escape from its parent association: e.g,, dynamical ejection and cluster outflows \citep{Pov67, Pfl10}.

Theoretical estimates for post-SN kick velocities are $v_{k}\sim$70--120\,km\,s$^{-1}$ \citep{Bra95}, assuming a non-eccentric, 1--25-d orbital period binary system composed of a 5\,$M_{\odot}$ star that explodes in an asymmetric supernova leaving behind a 1.4\,$M_{\odot}$ NS that tugs its hefty 15\,$M_{\odot}$ donor star. Velocities typical of isolated pulsars (200\,km\,s$^{-1}$ or even 350\,km\,s$^{-1}$) have been suggested depending on the initial conditions in the binary \citep{Por95,Van97}. \citet{Van00} distinguish between kick velocities in SGXB systems from those of BEXBs: $\sim$40 and $\sim$15\,km\,s$^{-1}$, respectively. Selecting only the 38 SGXBs or only the 35 BEXBs from the HMXB sample yield consistent results: both populations show $\gtrsim$6-$\sigma$ clustering with OB associations with no statistically significant differences between SGXBs and BEXBs.

In certain cases, the value of the kick velocity can be derived from the the radial velocity or proper motion when the identity of its parent OB association and the line-of-sight distances are known. The HMXB can then be traced back to its birthplace leading to an estimation of its kinematical age. Observations of the SGXB \object{LS\,5039} set $v\sim150$\,km\,s$^{-1}$ which constrains the time since its supernova to within 1.1\,Myr \citep{Rib02}. Another BEXB, \object{LS\,I\,$+$61$^{\circ}$303}, has $v = 27$\,km\,s$^{-1}$ and experienced its supernova $1.7\pm0.7$\,Myr ago \citep{Mir04}. The SGXB \object{4U\,1700$-$377} shows $v= 75$\,km\,s$^{-1}$ with its supernova $2.0\pm0.5$\,Myr ago \citep{Ank01}. \object{Vela\,X-1}, which is also a SGXB, has $v= 90$\,km\,s$^{-1}$ with $2\pm1$\,Myr having elapsed since its supernova \citep{Van96}. Recently, the velocity of the SGXB 4U\,1907$+$09 was measured at $\sim160$\,km\,s$^{-1}$ with respect to the Galactic Plane. Its kinematical age is only 0.1\,Myr \citep{Gva11}. The latter two systems present bow shocks as they move through the interstellar medium confirming the supersonic velocities of these HMXBs. Table\,\ref{tab_kick} summarizes these measurements.

We emphasize that there is no clearly-identified OB association for most HMXBs in our sample. Yet the vestiges of this dynamic history are imprinted on the overall HMXB distribution. Consider the distance separating each HMXB from its nearest OB association \citep[e.g.,][]{Van00}. Excluding separations larger than 1\,kpc, because they are less likely to be from post-natal kicks, we obtain an average minimum distance of $r_{\mathrm{min}}=$0.4$\pm$0.2\,kpc. With this $r_{\mathrm{min}}$, and with $\xi$ maximized at $\tau \sim -4$\,Myr (Fig.\,\ref{xi_hmxb_rot}), we estimate an average velocity of 100$\pm$50\,km\,s$^{-1}$ for the HMXBs in our sample. This is consistent with measurements of the kick velocities in individual objects \citep[see Table\,\ref{tab_kick} and, e.g., ][]{Gun70,Sto79}.

Alternatively, with $r_{\mathrm{min}}=0.4\pm0.2$\,kpc and assuming $v = 100$\,km\,s$^{-1}$, this translates to a migration timescale of 4$\pm$2\,Myr which is consistent with theoretical predictions for the average kinematical age of runaway massive binaries \citep{Bla61,Sto79,Van89}. Notice that these kinematical age and distance scales are consistent with the distribution of $\xi$ whose value is maximized between $-2$ and $-7$\,Myr (Fig.\,\ref{xi_hmxb_rot}). The lack of an equivalent increase in the value of $\xi$ between $+2$ and $+7\,Myr$ (which we would expect for kicks in random directions) is consistent with the effects of Galactic rotation, and is supported by \citet{Car05} who found a significant deficiency in the number of objects moving in retrograde Galactic orbits, at least for metal-poor binary stars in the Solar neighborhood. The range of shifts that increase the correlation amplitude is wide reflecting the broad parameter space of velocities and kinematical ages represented within the HMXB class. Dynamical ejection from the cluster prior to the supernova phase can lead to runaway velocities of the order of 150--200\,km\,s$^{-1}$ \citep{Pov67,Gie86,Pfl10}. This would make it difficult to retrace the trajectory of the HMXB back to its parent association, and the migration distance would be larger than expected from a kick alone. This is another factor contributing to the wide range in the shifts that increase the amplitude of $\xi$. 

Thus, the observed distribution of HMXBs in the Milky Way is consistent with the view that these systems have high velocities, on average. \citet{1998AaA...330..201C} arrived at the same conclusion by deriving the peculiar velocities of 17 HMXBs using \emph{HIPPARCOS} observations (see also \citet{Van00} concerning unreliable distance estimates for these sources). This feature is not unique to the Milky Way since HMXBs in the Small Magellanic Cloud also possess large velocities, on average \citep{Coe05}.

Minimum separation distances can provide clues to the HMXB-OB connection in specific cases \citep[e.g.,][]{Van00}. For example, there are 8 HMXBs in our sample for which the uncertainty on the line-of-sight distance is smaller than the distance separating it from its nearest OB association: \object{1H~1249$-$637}, \object{Cyg~X-3}, \object{EXO~2030$+$375}, \object{gam~Cas}, \object{IGR~J18027$-$2016}, \object{IGR~J18483$-$0311}, \object{IGR~J19140$+$0951}, and \object{SS~433}. Those with separations smaller than the average minimum separation distance ($\sim$0.4\,kpc) are prime examples of HMXBs whose migration distances are consistent with those expected from post-natal kicks. Small separation distances can result from kick velocities that are lower than average, which can occur during the formation of a black hole since less material is expected to be expelled during the supernova (or gamma-ray burst) that created it \citep[e.g.,][]{Mir03}.

A separation distance larger than 0.4\,kpc could imply that the binary inherited a higher kick velocity than average, or that it experienced dynamical ejection prior to its supernova (placing the system further away from its OB association than would be expected from its velocity and kinematical age). Another possibility is that the OB association that produced the HMXB is no longer active enough to be catalogued: massive stars can photo-evaporate their molecular clouds on timescales of 30\,Myr \citep[e.g.,][]{Wil97}.

On the other hand, if the HMXB distance is relatively accurate (i.e., within $\pm$0.5\,kpc), and if the distance to the nearest OB association is less than this uncertainty, then this increases the likelihood that the objects were linked in the past. This is the case for 12 HMXBs in our sample: \object{4U~1700$-$377}, \object{Cyg~X-1}, \object{GX~301$-$2}, \object{GX~304$-$1}, \object{H~1145$-$619}, \object{IGR~J01583$+$6713}, \object{RX~J0146.9$+$6121}, \object{RX~J0440.9$+$4431}, \object{RX~J1826.2$-$1450}, \object{SAX~J1818.6$-$1703}, \object{Vela~X-1}, and \object{X~Per}.

Table\,\ref{tab_HMXB_OB_sep} compiles these results. This assumes (perhaps unjustifiably) that the nearest OB association is the most likely birthplace of the HMXB being considered. Confirmation of the link (or lack thereof) between an HMXB and the nearest OB association is only possible through a detailed study of the velocities of the objects, which is beyond the scope of this paper. 

We close this section by pointing out that of the \numHMXB\ HMXBs in our sample, 9 are located more than 1\,kpc from their nearest OB association. Thus, the vast majority (89\%) of the HMXBs in our sample are situated close ($\lesssim$1\,kpc) to an OB association. This is consistent with \citet{Zin07} who estimate that 5--10\% of all OB stars form away from any cluster or association. Distance uncertainties are reported for 6 of 9 such sources, and these uncertainties exceed 1.5\,kpc, except for \object{H~0115$+$634} which has an uncertainty of $\pm$1\,kpc. This BEXB is situated more than 2\,kpc from its nearest OB association. This could represent a case where the young stellar binary was formed in isolation \citep{Zin07}.

\section{Summary \& Conclusions}
\label{conc}

This is the first time that the clustering of HMXBs and OB associations in the Milky Way has been demonstrated statistically. Since the correlation function relates objects in Cartesian space (adding a third dimension is trivial), this is a more robust spatial relation than those derived in prior studies which focused on the distribution of a single dimension such as longitudes or galactocentric radii. 

Not only does the correlation function confirm the expected view that HMXBs and OB associations are clustered together, the characteristic scale of the correlation contains the vestiges of stellar and Galactic evolution. Migration due to the kick velocity gained by the binary after the supernova can be constrained to a few hundred parsecs. This translates to kinematical ages (time spanning the supernova and HMXB phases) of $\sim$4\,Myr, which is consistent with theoretical expectations. We point out that these results are based on the distributions of the HMXB and OB populations which compose the ``grand design'' of the Milky Way. Yet even with this global view, the correlation function allows us to deduce the effects of local perturbations such as kick velocities. 

The correlation function opens novel areas of research. Our upcoming \emph{Chandra} survey of the Norma Arm will help locate dozens of new HMXBs which will allow us to examine the clustering of HMXBs and OB associations within a specific arm. The structure of these spiral arms should come into sharper focus with the \emph{Gaia} mission \citep{Per01}. The next hard X-ray surveyor, \emph{NuSTAR} \citep{Har10b}, will provide unprecedented sensitivity at these energies in the hunt for HMXBs. An increase in the discovery space of HMXB populations should allow us to probe deeper into the evolutionary history of massive stars and compact objects. This will permit a better understanding of the stellar content and its distribution in the Galaxy.

\acknowledgments
The authors warmly thank the referee for useful discussions that led to an improved manuscript. AB thanks Prof. I.F. Mirabel for discussions on the nature of HMXBs, and Dr. N. Barri\`{e}re for help with IDL. AB and JT acknowledge partial support from \chan\ award number G08-9055X issued by the \chan\ X-ray Observatory Center, which is operated by the Smithsonian Astrophysical Observatory for and on behalf of the National Aeronautics and Space Administration (NASA), under contract NAS8-03060. This research has made use of: data obtained from the High Energy Astrophysics Science Archive Research Center (HEASARC) provided by NASA's Goddard Space Flight Center; the SIMBAD database operated at CDS, Strasbourg, France; NASA's Astrophysics Data System Bibliographic Services; and the IGR Sources page (\texttt{http://irfu.cea.fr/Sap/IGR-Sources}).

\bibliographystyle{apj}
\bibliography{bod.bib}

\pagestyle{empty}
\begin{deluxetable}{ l r r c l l }
\tablenum{1}
\tablewidth{0pt}
\tablecaption{Distances reported for the high-mass X-ray binaries used in this analysis}
\tablehead{
\colhead{Name}           & \colhead{$l$}      &
\colhead{$b$}          & \colhead{distance [kpc]}  &
\colhead{classification}    &  \colhead{reference}  }
\startdata

1A 0535$+$262		& 181.445        & $-$2.644	       & $2_{-0.7}^{+0.4}$	       & BEXB (P, QPO, C, T)         & \citet{1998MNRAS.297L...5S}  \\
1A 1118$-$615			& 292.500        & $-$0.892	       & $5\pm2$		       & BEXB (P, QPO, C, T)         & \citet{1981AaA....99..274J}  \\
1E 1145.1$-$6141		& 295.490        & $-$0.010	       & $8.5\pm1.5$	       & SGXB (P)		       & \citet{2002ApJ...581.1293R}  \\
1H 1249$-$637			& 301.958        & $-$0.203	       & $0.392\pm0.055$       & BEXB (P?)  	       & \citet{2009AaA...507..833M}  \\
3A 0114$+$650			& 125.710        & $+$2.563	       & $4.5\pm1.5$	       & HMXB (P, E)	       & \citet{2008MNRAS.389..608F}  \\
3A 0656$-$072			& 220.128        & $-$1.769  	       & $3.9\pm0.1$	       & BEXB (P, C, T)	       & \citet{2006AaA...451..267M}  \\
3A 1845$-$024			& 30.421         & $-$0.406	       & $10$		       & BEXB (P, T)	       & \citet{2002AaA...391..923G}  \\
3A 2206$+$543		& 100.603        & $-$1.106	       & $2.6$		       & SGXB (NS, QPO)	       & \citet{2006AaA...446.1095B}  \\
4U 1036$-$56			& 285.350        & $+$1.431	       & $5$		       & BEXB (P, T)	       & \citet{2002AaA...391..923G}  \\
4U 1700$-$377			& 347.754        & $+$2.173	       & $2.120\pm0.343$	       & SGXB		       & \citet{2009AaA...507..833M}  \\
4U 1909$+$07			& 41.896         & $-$0.811	       & $7\pm3$		       & SGXB (P)		       & \citet{2005MNRAS.356..665M}  \\
AX J1820.5$-$1434		& 16.473         & $+$0.068	       & $8.2\pm3.5$	       & BEXB (P, T)	       & \citet{1998ApJ...495..435K}  \\
Cen X-3				& 292.091        & $+$0.337	       & $10\pm1$		       & SGXB (P, E, QPO, C, VHE)    & \citet{1979ApJ...229.1079H}  \\
Cyg X-1				& 71.335         & $+$3.067	       & $2.1\pm0.25$	       & SGXB (BHC, muQSO)	       & \citet{2005MNRAS.358..851Z}  \\
Cyg X-3				& 79.845         & $+$0.700	       & $7.2_{-0.5}^{+0.3}$		       & SGXB (BHC, muQSO, QPO)      & \citet{2009ApJ...695.1111L}  \\
EXO 0331$+$530		& 146.052        & $-$2.194	       & $7.5\pm1.5$	       & BEXB (P, QPO, C, T)         & \citet{1999MNRAS.307..695N}  \\
EXO 2030$+$375		& 77.152         & $-$1.242	       & $7.1\pm0.2$	       & BEXB (P, QPO, T)	       & \citet{2002ApJ...570..287W}  \\
gam Cas				& 123.577        & $-$2.148	       & $0.117\pm0.012$	       & BEXB		       & \citet{2009AaA...507..833M}  \\
Ginga 1843$+$009		& 33.038         & $+$1.690	       & $12.5\pm2.5$	       & BEXB (P, T)	       & \citet{2001AaA...371.1018I}  \\
GRO J1008$-$57		& 283.000        & $-$1.822	       & $5$		       & BEXB (P, C, T)	       & \citet{2002AaA...391..923G}  \\
GT 0236$+$610		& 135.675        & $+$1.086	       & $2.5$		       & BEXB (BHC, muQSO, VHE)      & \citet{2002AaA...391..923G}  \\
GX 301$-$2			& 300.098        & $-$0.035	       & $3.5\pm0.5$		       & HMXB (P, C, T)	       & \citet{2006AaA...457..595K}  \\
GX 304$-$1			& 304.103        & $+$1.247	       & $2.4\pm0.5$	       & BEXB (T)		       & \citet{1980MNRAS.190..537P}  \\
H 0115$+$634			& 125.924        & $+$1.026	       & $8\pm1$		       & BEXB (P, QPO, B, C, T)      & \citet{2001AaA...369..108N}  \\
H 1145$-$619			& 295.611        & $-$0.240	       & $3.1\pm0.5$	       & BEXB (P, T)	       & \citet{1997MNRAS.288..988S}  \\
H 1417$-$624			& 313.021        & $-$1.598	       & $6\pm5$		       & BEXB (P, T)	       & \citet{1984ApJ...276..621G}  \\
H 1538$-$522			& 327.419        & $+$2.164	       & $4.5$  	       & SGXB (P, E, C)	       & \citet{2004ApJ...610..956C}  \\
H 1907$+$097			& 43.744         & $+$0.476	       & $5$		       & SGXB (P, C, T)	       & \citet{2005AaA...436..661C}  \\
IGR J00370$+$6122		& 121.221        & $-$1.464	       & $3$		       & SGXB (P)		       & \citet{2004ATel..285....1N}  \\
IGR J01363$+$6610		& 127.394        & $+$3.726	       & $2$		       & BEXB (T)		       & \citet{2005AaA...440..637R}  \\
IGR J01583$+$6713		& 129.352        & $+$5.189	       & $4\pm0.4$  	       & BEXB (P, T)	       & \citet{2008MNRAS.386.2253K}  \\
IGR J06074$+$2205		& 188.385        & $+$0.814	       & $1$		       & BEXB		       & \citet{2006AaA...455...11M}  \\
IGR J08262$-$3736		& 256.425        & $+$0.296         & $6.1$		       & SGXB		       & \citet{2010AaA...519A..96M}  \\
IGR J08408$-$4503		& 264.040        & $-$1.952	       & $2.7$		       & SFXT		       & \citet{2007AaA...465L..35L}  \\
IGR J11215$-$5952		& 291.893        & $+$1.073	       & $6.2$		       & SFXT (P)		       & \citet{2006AaA...449.1139M}  \\
IGR J11305$-$6256		& 293.945        & $-$1.485	       & $3$		       & BEXB (T)		       & \citet{2006AaA...449.1139M}  \\
IGR J11435$-$6109		& 294.881        & $+$0.687	       & $8.6$		       & BEXB (P, T)	       & \citet{2009AaA...495..121M}  \\
IGR J13020$-$6359		& 304.089        & $-$1.120	       & $5.5\pm1.5$	       & BEXB (P, T)	       & \citet{2005MNRAS.364..455C}  \\
IGR J14331$-$6112		& 314.846        & $-$0.764	       & $10$		       & HMXB		       & \citet{2008AaA...482..113M}  \\
IGR J16195$-$4945		& 333.557        & $+$0.339	       & $4.5\pm2.5$	       & SFXT		       & \citet{2005AaA...429L..47S}  \\
IGR J16207$-$5129		& 332.459        & $-$1.050	       & $6.1_{-3.5}^{+8.9}$         & SGXB (E?)  	       & \citet{2008AaA...486..911N}  \\
IGR J16318$-$4848		& 335.617        & $-$0.448	       & $1.6$		       & HMXB (sgB[e])	       & \citet{2008AaA...484..801R}  \\
IGR J16320$-$4751		& 336.330        & $+$0.169	       & $3.5$		       & SGXB (P)		       & \citet{2008AaA...484..801R}  \\
IGR J16327$-$4940		& 335.093        & $-$1.132	       & $2$		       & SGXB (T)		       & \citet{2010AaA...519A..96M}  \\
IGR J16358$-$4726		& 337.099        & $-$0.007	       & $7\pm1$		       & HMXB (P, Be?)        & \citet{2006AaA...455.1165L}  \\
IGR J16393$-$4643		& 338.002        & $+$0.075	       & $10.6$		       & SGXB (P)		       & \citet{2008AaA...484..783C}  \\
IGR J16418$-$4532		& 339.189        & $+$0.489	       & $13$		       & SGXB (P, E, SFXT?)	       & \citet{2008AaA...484..783C}  \\
IGR J16465$-$4507		& 340.054        & $+$0.135	       & $9.5_{-5.7}^{+14.1}$        & SFXT (P, E?)	       & \citet{2008AaA...486..911N}  \\
IGR J16479$-$4514		& 340.163        & $-$0.124	       & $2.8_{-1.7}^{+4.9}$         & SFXT (E, P?)	       & \citet{2008AaA...486..911N}  \\
IGR J17252$-$3616		& 351.497        & $-$0.354	       & $8_{-2}^{+2.5}$	       & SGXB (P, E)	       & \citet{2009AaA...505..281M}  \\
IGR J17391$-$3021		& 358.068        & $+$0.445	       & $2.7$		       & SFXT		       & \citet{2008AaA...484..801R}  \\
IGR J17404$-$3655		& 352.626        & $-$3.273	       & $9.1$		       & HMXB		       & \citet{2009AaA...495..121M}  \\
IGR J17544$-$2619		& 3.236          & $-$0.336	       & $3.2\pm1$  	       & SFXT (NS?) 	       & \citet{2006AaA...455..653P}  \\
IGR J18027$-$2016		& 9.420          & $+$1.036	       & $12.4\pm0.1$	       & SGXB (P, E)	       & \citet{2010AaA...510A..61T}  \\
IGR J18214$-$1318		& 17.681         & $+$0.485	       & $8\pm2$		       & SGXB		       & \citet{2009ApJ...698..502B}  \\
IGR J18406$-$0539		& 26.670         & $-$0.173	       & $1.1$		       & BEXB (muQSO?)	       & \citet{2006AaA...448..547M}  \\
IGR J18410$-$0535		& 26.764         & $-$0.239	       & $3.2_{-1.5}^{+2}$	       & SFXT (P)		       & \citet{2008AaA...486..911N}  \\
IGR J18450$-$0435		& 28.138         & $-$0.657	       & $3.6$		       & SFXT		       & \citet{1996MNRAS.281..333C}  \\
IGR J18483$-$0311		& 29.750         & $-$0.745	       & $2.83\pm0.05$	       & SFXT (P)		       & \citet{2010AaA...510A..61T}  \\
IGR J19113$+$1533		& 49.010         & $+$2.758	       & $9.1$		       & HMXB (sgB[e], T)	       & \citet{2010AaA...519A..96M}  \\
IGR J19140$+$0951		& 44.296         & $-$0.469	       & $3.6\pm0.04$	       & SGXB (NS?) 	       & \citet{2010AaA...510A..61T}  \\
IGR J21347$+$4737		& 92.172         & $-$3.117	       & $5.8$		       & HMXB (T, SFXT?, Be?)        & \citet{2009AaA...495..121M}  \\
KS 1947$+$300		& 66.099         & $+$2.083	       & $9.5\pm1.1$	       & BEXB (P, T)	       & \citet{2005AstL...31...88T}  \\
OAO 1657$-$415		& 344.369       & $+$0.319	& $7.1\pm1.3$	       & SGXB (P, E)	       & \citet{2006MNRAS.367.1147A}  \\
PSR B1259$-$63		& 304.165       & $-$1.011	       & $2.75_{-0.49}^{+0.61}$		       & BEXB (P, VHE)	       & \citet{2002astro.ph..7156C}  \\
RX J0146.9$+$6121		& 129.541       & $-$0.800	       & $2.3\pm0.5$	       & BEXB (P, T?)	       & \citet{1997AaA...322..183R}  \\
RX J0440.9$+$4431		& 159.847       & $-$1.270	       & $3.3\pm0.5$	       & BEXB (P)		       & \citet{2005AaA...440.1079R}  \\
RX J1826.2$-$1450		& 16.882         & $-$1.289	       & $2.5\pm0.1$	       & SGXB (BHC, muQSO, VHE)      & \citet{2005MNRAS.364..899C}  \\
SAX J1818.6$-$1703	& 14.080         & $-$0.704	       & $2.1\pm0.1$	       & SFXT		       & \citet{2010AaA...510A..61T}  \\
SAX J2103.5$+$4545	& 87.130         & $-$0.685	       & $6.8\pm2.3$	       & BEXB (P, QPO, T)	       & \citet{2010MNRAS.401...55R}  \\
SS 433				& 39.694         & $-$2.245	       & $5.5\pm0.2$	       & SGXB (BHC, muQSO)	       & \citet{2004ApJ...616L.159B}  \\
SWIFT J2000.6$+$3210	& 68.986         & $+$1.134	       & $8$		       & BEXB (P, T)	       & \citet{2008AaA...482..113M}  \\
Vela X-1				& 263.058       & $+$3.930  	& $1.4\pm0.5$	       & SGXB (P, E, C)	       & \citet{1998AaA...330..201C}  \\
X Per				& 163.081       & $-$17.136	& $0.801\pm0.138$	       & BEXB (P, C)	       & \citet{2009AaA...507..833M}  \\
XTE J1543$-$568		& 324.955       & $-$1.475	       & $10$		       & BEXB (P, T)	       & \citet{2002AaA...391..923G}  \\
XTE J1810$-$189		& 11.362         & $+$0.059	       & $11.5$		       & HMXB (NS, T)	       & \citet{2008ATel.1443....1M}  \\
XTE J1829$-$098		& 21.697         & $+$0.279	       & $10$		       & HMXB (P, T, Be?)	       & \citet{2007ApJ...669..579H}  \\
XTE J1855$-$026		& 31.076         & $-$2.092	       & $10$		       & SGXB (P, E, T)	       & \citet{2002AaA...391..923G}  \\
XTE J1946$+$274		& 63.207         & $+$1.396	       & $9.5\pm2.9$	       & BEXB (P, C, T)	       & \citet{2003ApJ...584..996W}

\enddata
\vspace{-8mm}
\tablecomments{BEXB=Be X-ray binary; BHC=black hole candidate; C=cyclotron; E=eclipsing; HMXB=high-mass X-ray binary;\vspace{-1mm} muQSO=microquasar; NS=neutron star; P=pulsations; QPO=quasi-periodic oscillations; SFXT=supergiant fast X-ray transient;\vspace{-1mm} sgB[e]=supergiant B-emission line star; SGXB=supergiant X-ray binary; T=transient; VHE=very high energy}
\label{tab_hmxb}
\end{deluxetable}

\begin{deluxetable}{ l c c c l }
\tablenum{2}
\tablewidth{0pt}
\tablecaption{Kick velocities and kinematical ages reported for a few well-known HMXBs}
\tablehead{
\colhead{name}           & \colhead{velocity [km\,s$^{-1}$]}  &
\colhead{kinematical age [Myr]}    & \colhead{migration distance [kpc]} & \colhead{reference}  }
\startdata

\object{4U\,1700$-$377}			& 75		& 2.0		& 0.2 	& \citet{Ank01}  \\

\object{LS\,5039}				& 150	& 1.1 	& 0.2	 	& \citet{Rib02}  \\

\object{LS\,I\,$+$61$^{\circ}$303}	& 27		& 1.7		& 0.05 	& \citet{Mir04}  \\

\object{Vela\,X-1}				& 160	& 0.1		& 0.02 	& \citet{Gva11} \\	

\enddata
\label{tab_kick}
\end{deluxetable}

\begin{deluxetable}{ l c c }
\tablenum{3}
\tablewidth{0pt}
\tablecaption{Minimum separation distances between HMXBs and OB associations in selected cases}
\tablehead{
\colhead{name}           & \colhead{HMXB distance uncertainty [kpc]}  &
\colhead{distance to nearest OB assoc. [kpc]}  }
\startdata
\cutinhead{HMXB distance uncertainty $<$ distance to nearest OB assoc.}
\object{1H~1249$-$637} 		& 0.06	& 0.1 	\\
\object{Cyg~X-3} 			& 0.5 	& 0.7 	\\
\object{EXO~2030$+$375} 	& 0.2 	& 0.6 	\\
\object{gam~Cas} 			& 0.02 	& 0.08 	\\
\object{IGR~J18027$-$2016} 	& 0.1 	& 0.8 	\\
\object{IGR~J18483$-$0311} 	& 0.05 	& 0.2 	\\
\object{IGR~J19140$+$0951} 	& 0.04 	& 0.3 	\\
\object{SS~433} 			& 0.2 	& 0.7 	\\

\cutinhead{HMXB distance uncertainty $\ge$ distance to nearest OB assoc.} 
\object{4U~1700$-$377} 		& 0.5 	& 0.2 	\\
\object{Cyg~X-1} 			& 0.25 	& 0.2 	\\
\object{GX~301$-$2} 		& 0.5 	& 0.2 	\\
\object{GX~304$-$1} 		& 0.5 	& 0.1 	\\
\object{H~1145$-$619} 		& 0.5 	& 0.2 	\\
\object{IGR~J01583$+$6713} 	& 0.4 	& 0.4 	\\
\object{RX~J0440.9$+$4431} 	& 0.5 	& 0.5		\\
\object{RX~J0146.9$+$6121} 	& 0.5 	& 0.4 	\\
\object{RX~J1826.2$-$1450} 	& 0.1 	& 0.004 	\\
\object{SAX~J1818.6$-$1703} 	& 0.1	 	& 0.1 	\\
\object{Vela~X-1} 			& 0.5 	& 0.03	\\
\object{X~Per} 				& 0.3 	& 0.1		\\
\enddata
\label{tab_HMXB_OB_sep}
\end{deluxetable}

\end{document}